\documentclass[%
reprint,
aps,
prl,
showpacs,
superscriptaddress,
nofootinbib,
nobibnotes,
amsmath,amssymb,
]{revtex4-1}
\usepackage{graphicx}
\usepackage[utf8]{inputenc}
\usepackage[T1]{fontenc}
\usepackage{lmodern}
\usepackage[english]{babel}
\usepackage{siunitx}
\usepackage{hyperref}

\DeclareSIUnit\molar{\mole\per\cubic\deci\metre}
\DeclareSIUnit\Molar{M}
\DeclareSIUnit\fps{fps}

\begin{document}

\title{Tuning Synthetic Semiflexible Networks by Bending Stiffness}
\thanks{This article was accepted for publication in \href{http://journals.aps.org/prl/}{Physical Review Letters} published by the American Physical Society.}

\author{Carsten Schuldt}
\thanks{C.S. and J.S. contributed equally to this work.}
\author{Jörg Schnauß}
\thanks{C.S. and J.S. contributed equally to this work.}
\affiliation{Institute of Experimental Physics I, Universität Leipzig, Linnéstraße 5, 04103 Leipzig, Germany}
\affiliation{Fraunhofer Institute for Cell Therapy and Immunology, Perlickstraße 1, 04103 Leipzig, Germany}
\author{Tina Händler}
\author{Martin Glaser}
\affiliation{Institute of Experimental Physics I, Universität Leipzig, Linnéstraße 5, 04103 Leipzig, Germany}
\affiliation{Fraunhofer Institute for Cell Therapy and Immunology, Perlickstraße 1, 04103 Leipzig, Germany}
\author{Jessica Lorenz}
\affiliation{Fraunhofer Institute for Cell Therapy and Immunology, Perlickstraße 1, 04103 Leipzig, Germany}
\author{Tom Golde}
\author{Josef A. Käs}
\affiliation{Institute of Experimental Physics I, Universität Leipzig, Linnéstraße 5, 04103 Leipzig, Germany}
\author{David M. Smith}
\affiliation{Fraunhofer Institute for Cell Therapy and Immunology, Perlickstraße 1, 04103 Leipzig, Germany}
\date{\today}

\begin{abstract}
\label{sec-1}

The mechanics of complex soft matter often cannot be understood in the classical physical frame of flexible polymers or rigid rods.
The underlying constituents  are semiflexible polymers, whose finite bending stiffness ($\kappa$) leads to non-trivial mechanical responses.
A natural model for such polymers is the protein actin.
Experimental studies of actin networks, however, are limited since the persistence length ($l_{\text{p}} \propto \kappa$) cannot be tuned.
Here, we experimentally characterize this parameter for the first time in entangled networks formed by synthetically produced, structurally tunable DNA nanotubes.
This material enabled the validation of characteristics inherent to semiflexible polymers and networks thereof, i.e., persistence length, inextensibility, reptation and mesh size scaling.
While the scaling of the elastic plateau modulus with concentration \(G_0 \propto c^{7/5}\) is consistent with previous measurements  and established theories, the emerging persistence length scaling \(G_0 \propto l_{\text{p}}\) opposes predominant theoretical predictions.

\end{abstract}


\maketitle
\label{sec-2}

Semiflexible polymers are of fundamental importance to biological systems due to their ability to form stable scaffolds at low volume fractions, with their voids providing free space for the molecular transport and metabolic processes necessary for active, living matter \cite{huber_emergent_2013,broedersz_modeling_2014}.

This special class of polymers is distinguished by their non-vanishing bending rigidity, featuring an outstretched configuration while still subjected to strong thermal fluctuations \cite{huber_emergent_2013}.
Within the worm-like chain model \cite{kratky_rontgenuntersuchung_1949,saito_statistical_1967}, this stiffness is described by the persistence length (\(l_{\text{p}}\)), which represents the decay constant of the tangent-tangent correlation \cite{doi_theory_1986}.
When the contour length (\(l_{\text{c}}\)) is comparable to \(l_{\text{p}}\), polymers are considered semiflexible \cite{huber_emergent_2013}.
Networks of these filamentous constituents in the entangled concentration regime (mesh size \(\xi \leq l_{\text{p}}\)) have been subject to considerable studies \cite{mueller_viscoelastic_1991,mackintosh_elasticity_1995,gardel_elastic_2004,fletcher_cell_2010,sonn-segev_viscoelastic_2014}, but their central quantity \(l_{\text{p}}\) is still unexplored since it remained experimentally inaccessible to systematic variation for any given material. 
Sophisticated theoretical approaches were built starting from the single-filament level in order to describe the emergence of the non-trivial mechanical behaviors since classical models for flexible chains or rigid rods are not directly applicable \cite{broedersz_modeling_2014}.
Depending on the microscopic models for network architecture and load transduction to the individual filament, markedly different scaling predictions for the linear elastic plateau shear modulus (\(G_0\)) with respect to concentration ($c$) and $l_{\text{p}}$ have been suggested \cite{mackintosh_elasticity_1995,kroy_force-extension_1996,hinner_entanglement_1998,isambert_dynamics_1996}.

The concentration scaling is readily experimentally accessible, and corresponding theoretical predictions have been rigorously verified \cite{hinner_entanglement_1998,tassieri_dynamics_2008,liu_microrheology_2006,hinsch_non-affine_2009}.
In pioneering experimental work, the cellular biopolymer actin was established as a model system for semiflexible filaments and is still considered the gold standard \cite{kas_f-actin_1996,isambert_dynamics_1996,mueller_viscoelastic_1991,hinner_entanglement_1998,morse_viscoelasticity_1998-1,palmer_diffusing_1999,gardel_microrheology_2003,liu_microrheology_2006},  although some of its material-specific effects such as the unidirectional ``treadmilling'' of monomers from one end toward the other cannot be considered a general characteristic of semiflexible polymers \cite{huber_emergent_2013}.
However, the scaling of network elasticity with respect to \(l_{\text{p}}\) still remained experimentally inaccessible for this model system since the \(l_{\text{p}}\) of actin cannot be varied as an independent parameter \cite{tassieri_dynamics_2008}. 
This limitation is inherent for all biopolymers and thus a comprehensive validation of the published theoretical predictions is still pending to date.

To resolve this important dependency and to overcome the natural limitation imposed by the fixed \(l_{\text{p}}\) of actin, we employed synthetically produced, tile-based DNA tubes \cite{yin_programming_2008}, which were previously demonstrated to have tunable mechanical properties, i.e., stiffness \cite{schiffels_nanoscale_2013}.
They are soluble in water and stable for months in adequate pH and ionic conditions without displaying aging effects typical of  protein-based filaments \cite{yin_programming_2008}.
To assemble these tubes, we used a set of \(n\) ($4 \leq n \leq 14$) distinct, partially complementary DNA oligonucleotides (each 42 bases in length, see Supplemental Material sec. I), which hybridize to a half overlapping ring of \(n\) interconnected DNA helices (Fig. 1(a) \& (b)) \cite{yin_programming_2008}.
Axial sticky ends trigger a selective addition of matching oligonucleotides, inducing an effective polymerization of tubes with a contour length distribution comparable to actin \cite{yin_programming_2008}.
Depending on the set of \(n\) strands chosen, \(n\)-helix tubes (\(n\)HTs) with a uniform circumference were formed.
It was previously shown that the $l_{\text{p}}$ of these $n$HTs scales with their second moment of inertia \citep[see][]{schiffels_nanoscale_2013}.
Thus, these DNA $n$-helix tubes are purely synthetic polymers, programmable in their circumference and accordingly in their $l_{\text{p}}$ (Fig. 1(b)).

\begin{figure}
  \centering	
  \includegraphics{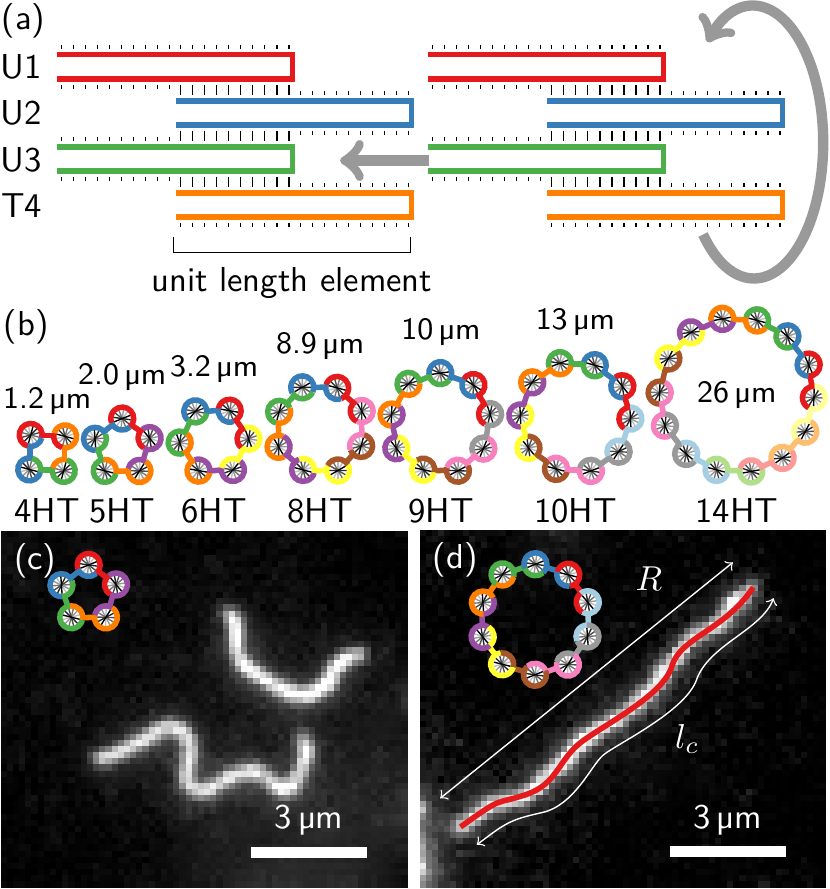}
  \caption{(Color online)
    \(n\)-Helix Tube (\(n\)HT) filament architecture.
    (a) Schematic of the assembly of a 4HT formed of four distinct 42-mers.
    Adjacent single-stranded DNA oligonucleotides share continuous complementary sections of 10 and 11 bases in length (long black ticks).
    Boundary strands U1 and T4 feature complementary sequences as well, enabling the formation of a tube-like ring from the planar sheet structure.
    The half-staggered motif of these rings with sticky ends on both sides promotes polymerization-like axial growth.
    (b) Cross sections of all seven different \(n\)HTs employed in this study with according measured values for \(l_p\).
    Strand numbers range from 4HT to 14HT and $l_{\text{p}}$ from \SIrange{1.2}{26}{\micro\meter}, respectively.
    (c \& d) Epi-fluorescent image of Cy3-labeled adsorbed \(n\)HTs.
    Left: 5HT.
    Right: 10HT.
    The red overlay is the filament contour approximated by image analysis with end-to-end distance $R$ and contour length $l_{\text{c}}$ \cite{smith_segmentation_2010}.
    a \& b inspired by Yin et al. \cite{yin_programming_2008}.}
	\label{fig:tube_design}
	\end{figure}

\label{sec-3}
\begin{figure}
 \centering	
\includegraphics{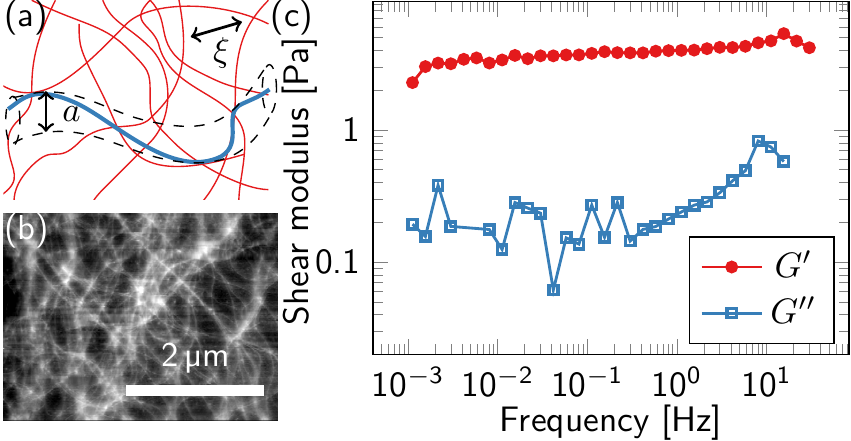}
\caption{(Color online)
  Network elasticity.
  (a) Schematic of a semiflexible filament meshwork with mesh size $\xi$.
  A test polymer (blue) of contour length $l_{\text{c}}$ is confined by the background of red filaments to a tube of width $a$.
  (b) AFM image of an 8HT network formed at \SI{4}{\micro\Molar}.
  (c) Frequency resolved viscoelastic properties of an 8HT network at \SI{6}{\micro\Molar} probed by bulk shear rheology at a strain of \SI{5}{\percent}.
  A pronounced elastic rubber plateau over more than four orders of magnitude is characteristic of all \(n\)HTs.
}
\label{fig:M1}
\end{figure}

\begin{figure*}
  \centering	
  \includegraphics{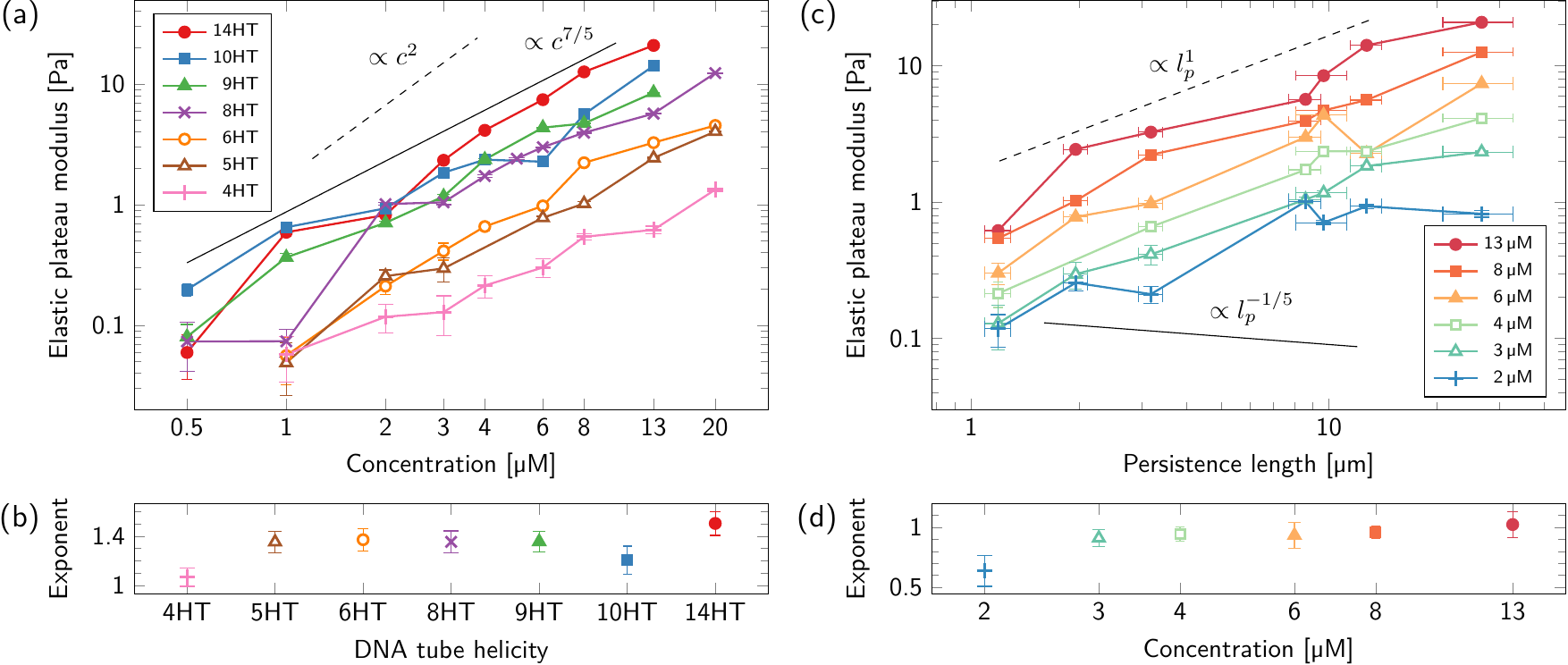}
  \caption{(Color online)
    $G_0$ scaling of $n$HT networks.
    (a) The concentration scaling for networks of different \(n\)HT is plotted.
    Predicted power laws from the tube model and unit cell approach are given by black and dashed lines, respectively.
    (b) Least-squares approximations yielded the actual power law exponents for all \(n\)HT, which accumulated around $\alpha \approx 7/5$.
    (c) The persistence length scaling for networks of different concentration is shown.
    (d) Approximated power law exponents yielded a mean exponent of \(\beta \approx 1 \).
    In summary, we found the elastic shear modulus to scale as $G_0 \propto c^{7/5}l_{\text{p}}$.
  } 
  \label{fig:central-result}
	\end{figure*}
We confirmed and extended upon previous measurements of $l_{\text{p}}$ that were based on freely fluctuating $n$HTs \cite{schiffels_nanoscale_2013} by evaluating epi-fluorescence images of more than 100 adsorbed, single filaments (Fig. 1(c) \& (d)) for each of the seven different $n$HTs used here.
One of the constituent oligonucleotides was substituted by a Cy3 conjugated analog for fluorescent imaging.
Image-based tracking and fitting of the tangent-tangent correlation yielded a range of \(l_{\text{p}}\) from \SI[separate-uncertainty]{1.2(1)}{\micro \meter} (4HT) to \SI[separate-uncertainty]{26(5)}{\micro \meter} (14HT).
Applying a kurtosis analysis (see Supplemental Material sec. II), we could exclude any possible effects from surface interactions during adsorption and found the same values and quantitative trend for $l_{\text{p}}$ as reported for freely fluctuating tubes \cite{schiffels_nanoscale_2013}.

While the determination of $l_{\text{p}}$ was performed in dilute samples to measure this intrinsic property, reptation experiments were performed in the same concentration regime as used for the rheology experiments.
A small number of fluorescently labeled tubes were embedded in an unlabeled network and their snake-like reptation motion was recorded and tracked as described for the $l_{\text{p}}$ determination (for more details see Supplemental Material sec. III).
 
Using this synthetic system, we were able to quantify the changes in emergent network mechanics resulting from systematically varying $l_{\text{p}}$ over more than one order of magnitude, without making any changes to the underlying material.
Each type of \(n\)HT forms isotropic networks (Fig. 2(a) \& (b)), which allowed us to investigate the mechanical response of these networks with dynamic shear rheology.
After an equilibration of \SI{2}{\hour}, a series of frequency ($f$) and strain ($\gamma$) sweeps was performed to characterize intrinsic mechanics as well as potential aging effects: (i) short $f$ sweep ($\gamma = \SI{5}{\percent}$, $f$ =  \SIrange{0.01}{30}{\hertz}, 5 data points per decade), (ii) long $f$ sweep ($\gamma = \SI{5}{\percent}$, $f$ = \SIrange{0.001}{30}{\hertz}, 21 data points per decade), (iii) short $f$ sweep, (iv) $\gamma$ sweep ($f= \SI{1}{\hertz}$, $\gamma$ = \SIrange{0.0125}{100}{\percent}, 20 data points per decade), (v) short $f$ sweep, and (vi) $\gamma$ sweep.
Different passivation techniques to inhibit air-liquid interface effects were also tested, yielding consistent results (cf. Supplemental Material sec. V).

Frequency sweeps revealed the predominately elastic response in the linear regime (Fig. 2(c)).
In the frequency range of \SIrange{e-3}{10}{\hertz}, the elastic modulus \(G'\) exceeded the viscous modulus \(G''\) by approximately one order of magnitude and no crossover between the two moduli was observed.
\(G'\) was nearly constant in this range and showed a broad rubber plateau over four decades.
This characteristic is in contrast to previous measurements on actin \cite{gardel_microrheology_2003}, where long-term relaxation and the associated viscoelastic crossover might be superimposed by treadmilling effects \cite{isambert_dynamics_1996,huber_emergent_2013} resulting from unbalanced binding kinetics of monomeric actin at filament ends.
In contrast, the \(n\)HTs employed here are not influenced by any type of treadmilling due to the high stability of hybridized DNA segments in the typical measurement conditions.
The dominance of $G’$, as well as the frequency independence of the rubber plateau, are universal features for all seven \(n\)HTs and all concentrations studied.
We derived the scaling of the elastic plateau modulus with respect to concentration from frequency sweeps with \(G_0 = G'(\SI{1}{\hertz},\SI{5}{\percent})\).
Between \SI{0.5}{\micro\Molar} and \SI{20}{\micro\Molar}, $G_0$ increased monotonically between \SI{50}{\milli\pascal} and \SI{20}{\pascal} for all \(n\)HTs (Fig. 3(a)).
Given a power law $G_0 \propto c^{\alpha}$, exponents accumulated around \(\alpha = 7/5\) (Fig. 3(b)).
4HTs at low concentrations are very soft (low sub Pascal range) falling below the sensitivity of the rheometer.

These measurements allowed us to evaluate the scaling of $G_0$ with respect to  $l_{\text{p}}$ (Fig. 3(c)).
Circumferences of the \(n\)HTs were translated into $l_{\text{p}}$ with values determined as described above.
Each curve in Fig. 3(c) represents one specific concentration.
With increasing $l_{\text{p}}$, we found a monotonic increase of $G_0$.
Given a power law $G_0 \propto l_{\text{p}}^{\beta}$, exponents were found to accumulate around \(\beta = 1\) (Fig. 3(d)).
These scaling results were independent of the choice of strain and strain rate (cf. Supplemental Material sec. VII).
In conclusion, these shear rheology measurements accessed the concentration and $l_{\text{p}}$ dependencies of \(G_0\) and revealed an overall scaling relation of
\begin{equation}
G_0 \propto c^{7/5}l_{\text{p}} .
\end{equation}\par

To validate the applicability of our findings to the conceptual framework of an entropically fluctuating, entangled network, we ensured that these DNA \(n\)HT networks were indeed only topologically entangled and not physically cross-linked.
In particular, we investigated if these networks exhibited strain stiffening and whether single filaments displayed snake-like reptation within the background network.
These measurements revealed no sign of strain stiffening (Fig. 4(a)), which is considered a general characteristic of cross-linked networks \cite{gardel_elastic_2004,storm_nonlinear_2005} (cf. Supplemental Material sec. IV).
Additionally, filaments were observed to reptate freely in tube-like regions within the networks (Fig. 4(b) and Supplemental Material sec. III) \cite{de_gennes_remarks_1976}.
This characteristic thermal motion for entangled networks would be suppressed by cross-links \cite{kas_direct_1994}.
Furthermore, spatio-temporal traces of reptating filaments revealed the dependence of the network mesh size on the monomer concentration was comparable to $\xi \propto c^{-1/2}$ (Fig. 4(b) inset).
This behavior was consistent with our own reptation-based data on actin filaments, as well as both the predicted \cite{de_gennes_remarks_1976} and reported scalings for networks of semiflexible polymers as measured by micro-particle diffusion \cite{schmidt_chain_1989}.
Throughout all different elasticities, the mesh size was at the micrometer scale and shear elasticity dominated viscosity, which could be freely programmed by altering the nanoscale architecture of the underlying synthetic filaments.
Due to the newly discovered linear $l_{\text{p}}$ scaling of $G_0$, the bulk mechanics of this network could be precisely tuned  over a broad range while keeping the mesh size constant, in contrast to the structural impact of simply increasing the material concentration.
Finally, the inextensible nature of DNA \(n\)HTs was demonstrated through analysis of reptation data; filament fluctuations in equilibrium revealed bending modes without thermally excited stretching modes (see Supplemental Material sec. IV).

\begin{figure}[t]
  \centering	
  \includegraphics{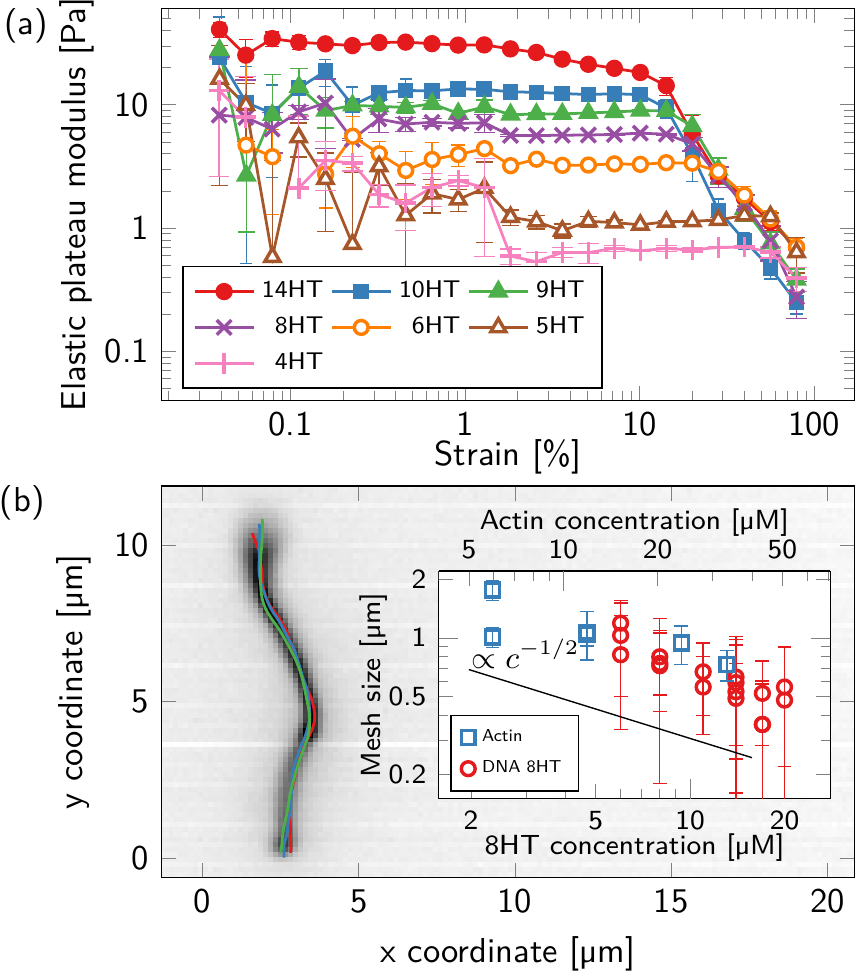}
  \caption{(Color online)
    Linearity and microscopic mesh size.
    (a) Strain sweeps of $G'(\SI{1}{\hertz})$ feature a broad linear elastic plateau for small strains.
    Beyond a characteristic strain $G'$ breaks off irreversibly depending on $n$HT.
    (b) Microscopic reptation of a single labeled DNA 8HT embedded in a network of the same type of filaments.
    The summation fluorescent image of 600 sampled configurations is overlaid by three exemplary filaments as found by image analysis ($c= \SI{14}{\micro \Molar}$).
    Inset: Mesh size of actin and DNA 8HT networks as derived from such reptation analysis at different concentrations.
    Monomeric concentrations are shifted by the chain  length conversion factor \(x = l_{\text{monomer,}n\text{HT}} / l_{\text{monomer, actin}} \simeq 2.5\) to compare the same concentrations of contour length per unit volume.
    Mesh sizes are on the same order of magnitude and decrease with concentration.
    Theoretical scaling prediction \(\xi \propto c^{-1/2}\) is indicated by the solid line \cite{de_gennes_remarks_1976}.
  }
  \label{fig:linearity-repatation}
	\end{figure}


Consequently, the DNA \(n\)HT networks can be considered purely entangled and compare well to the previously employed model system of actin networks.
The predominant theoretical description of entangled networks is the \emph{tube model} \cite{de_gennes_scaling_1979,morse_viscoelasticity_1998-1}.
Within this framework, the many-body problem of a network of filaments is reduced to a single fluctuating test-filament, with all other polymers conceptualized as an effective tube constricting its motions \cite{isambert_dynamics_1996,hinner_entanglement_1998,morse_tube_2001}.
Each collision with this tube contributes on the order of \(k_{\text{B}}T\) to the free energy of the system, with the sum of all collisions  constituting the shear modulus.
Shear deformations are modeled as a compression of this tube yielding the scaling \(G_0 \propto l_{\text{p}}^{-1/5} c_{\vphantom{p}}^{7/5} \), as also shown in numerical simulations \cite{hinsch_non-affine_2009}.
This trend has also emerged from different conceptual realizations of the interaction with the tube \cite{morse_tube_2001}.
Notably, these models predict a peculiar decrease of overall network elasticity as the individual components become stiffer, i.e., with increasing $l_{\text{p}}$, due  to the entropic origin of the mechanical response.
However, this important parameter was never comprehensively accessible to any previous experimental studies, and therefore never rigorously determined in an independent manner.

One attempt aimed to alter the ``effective'' persistence length of actin by varying solvent conditions \cite{tassieri_dynamics_2008}.
A persistence length scaling consistent with an effective medium implementation of the tube model was reported \cite{morse_tube_2001}, although only based on two independent $l_{\text{p}}$ values.
This clearly illustrates the natural limitations of biopolymers such as actin as comprehensive models systems for semiflexible filaments.
Intermediate filaments such as vimentin with a significantly lower $l_{\text{p}}$ also fall into this class and reveal typical semiflexible mechanical properties \cite{noding_intermediate_2012}.
However, both biopolymers not only differ in their mechanical properties but also in their underlying molecular structure and equilibrium-state dynamics, thereby rendering the consistent derivation of \(l_{\text{p}}\) scaling laws difficult.

Here, we present the first in-depth study on both parameters of concentration and persistence length performed without changing the underlying material.
Within our study, we corroborate the concentration scaling \(G_0 \propto c^{7/5}\) as predicted by the tube model \cite{morse_tube_2001}, which was already experimentally confirmed for actin networks \cite{hinner_entanglement_1998,tassieri_dynamics_2008} and more generally in numerical simulations \cite{hinsch_non-affine_2009}.
Though confirming several key characteristics of the tube model, we clearly find that $G_0$ scales linearly with $l_{\text{p}}$ -- in strong disagreement with the tube model's prediction $G_0 \propto l_{\text{p}}^{-1/5}$.
The overall scaling of the elastic modulus shown here of \(G_0 \propto c^{7/5} l_{\text{p}}\) cannot be explained with any existing theory.
The simplistic unit cell approach, for instance, predicts the linear \(l_{\text{p}}\) scaling correctly, but overestimates the impact of concentration \cite{satcher_jr_theoretical_1996,kroy_force-extension_1996}.
Another bending dominated theoretical approach is the so-called affine network model, which assumes local deformations, namely affine contraction and stretching of individual filaments.
This approach yields an elastic plateau modulus which increases with persistence length \( c_{\vphantom{p}}^{11/5} l_{\text{p}}^{7/5} \) \cite{mackintosh_elasticity_1995}, although it has been shown experimentally that this model most likely applies to cross-linked systems \cite{gardel_elastic_2004,jaspers_ultra-responsive_2014} and is inappropriate to describe the results of our $n$HT study.

Bending dominated theories, in particular the oversimplifying athermal unit cell model mentioned above \cite{kroy_force-extension_1996,satcher_jr_theoretical_1996}, predict the significantly stronger trend of $l_{\text{p}}$ scaling more closely than the tube model.
We therefore speculate that the tube model demands adjustment with respect to further internal energy contributions from filament bending, whose additional contributions scale linearly with filament stiffness \cite{satcher_jr_theoretical_1996}.

These findings contradict the established picture of semiflexible polymer networks, which are distinguished by the central quantity $l_{\text{p}}$.
Key aspects of the established tube model, such as reptation, inextensibility, and the correct concentration scaling were indeed proven for this \emph{de novo} model system of DNA $n$HTs.
In contrast, through rigorously studying the impact of $l_{\text{p}}$ for the first time, we found striking disagreement with the prediction of the tube model.
This observation is impossible with other known model systems, since they lacked the ability to tune \(l_{\text{p}}\) freely \cite{hinner_entanglement_1998} or deterministically \cite{fakhri_diameter-dependent_2009}, were cross-linked \cite{jaspers_ultra-responsive_2014}, were subjected to  treadmilling \cite{isambert_dynamics_1996,huber_emergent_2013}, or the ratio \(l_\text{c}/l_{\text{p}}\) did not fit within the semiflexible regime \cite{fakhri_diameter-dependent_2009}.
Due to the structurally modular, self-assembling nature of the DNA $n$HTs, other parameters of semiflexible networks such as the addition of physical cross-links might be readily accessible in an equally programmable way.
Since all of these material properties are decoupled for the first time, this unique model system will stimulate further development of our understanding of the emerging mechanics of semiflexible polymer networks.


\begin{acknowledgments}
We acknowledge funding by DFG (1116/17-1) and the Leipzig School of Natural Sciences “BuildMoNa” (GSC 185).
Part of this work has been supported through the Fraunhofer Attract project 601 683.
TH acknowledges funding by the European Social Fund (ESF—100077106).
We thank Angela Moore and Emilia Wisotzki for their help in editing the manuscript. We thank Klaus Kroy and Erwin Frey for fruitful discussions.
\end{acknowledgments}

\bibliography{newreferences}


\end{document}